\documentclass[useAMS,usenatbib,usegraphicx,referee]{mn2e}
\newcommand{\bugref}{\bibitem[\protect\citeauthoryear{dummy }{1893}]{dum}}

\title[Aligning VLBI Images of AGN]
{Aligning VLBI Images of Active Galactic Nuclei at Different Frequencies}

\author[Croke \& Gabuzda]{S. M. Croke$^{1,2}$ \& D. C. Gabuzda$^{3}$\\
$^{1}$Department of Physics, University of Stratyclyde,  
Glasgow G4 0NG, United Kingdom\\
$^{2}$Department of Mathematics, University of Glasgow, Glasgow, G12 8QW, 
United Kingdom\\
$^{3}$Department of Physics, University College Cork, Cork, Ireland} 

\begin{document}

\date{}
\pagerange{\pageref{firstpage}--\pageref{lastpage}} \pubyear{2007}
\maketitle
\label{firstpage}
\begin{abstract}
Many important techniques for investigating the properties of extragalactic
radio sources, such as spectral-index and rotation-measure mapping,
involve the comparison of images at two or more frequencies.  In the case
of radio interferometric data, this can be done by comparing the CLEAN maps 
obtained at the different frequencies. However, intrinsic differences in 
images due to the frequency dependence of the radio emission can be distorted by
additional differences that arise due to source
variability (if the data to be compared is obtained at different
times), image misalignment, and the frequency dependence of the
sensitivity to weak emission and the angular resolution provided by
the observations (the resolution of an interferometer depends
on the lengths of its baselines in units of the observing
wavelength).  These effects must be corrected for as best
as possible before multi-frequency data comparison techniques can
be applied.  We consider the origins for the
afore-mentioned factors, outline the standard
techniques used to overcome these difficulties, and describe in
detail a technique developed by us, based on the
cross-correlation technique widely used in other fields, to
correct for misalignments between maps at different frequencies.
\end{abstract}

\section{Introduction}

The radio emission of core-dominated, radio-loud Active Galactic Nuclei
(AGN) is synchrotron radiation generated in the relativistic jets that 
emerge from the nucleus of the galaxy, presumably along the rotational
axis of a central supermassive black hole. One important source of information 
about the physical conditions in the radio-emitting regions is the distribution
of the spectral index $\alpha$ over the source ($S_{\nu}\propto\nu^{\alpha}$, 
where $S_{\nu}$ is the source flux at frequency $\nu$). The core region
is typically observed to be at least partially optically thick, with a
nearly flat or inverted spectrum, while the jets are optically thin, with 
negative spectral indices. The spectrum may also flatten in regions of the jet 
in which there is re-acceleration of electrons or low-frequency absorption
(e.g. Gabuzda, Pushkarev \& Garnich 2000; Gabuzda, G\'omez \& Agudo 2001).
Synchrotron radiation can be
highly linearly polarised, to $\simeq 75\%$ in the case of a uniform magnetic
field (Pacholczyk 1970), and linear polarisation observations
can yield unique information about the orientation and degree of order of the
magnetic field in the synchrotron source, as well as the distribution of
thermal electrons and the magnetic-field geometry in the immediate vicinity
of the AGN (e.g., via Faraday rotation of the plane of polarisation).

The compact radio emission of such AGN can be probed with  high resolution
using Very Long Baseline Interferometry (VLBI). The radio telescopes
in the interferometric array can be separated by hundreds or thousands of
kilometres, making it infeasible to physically link (synchronise) them 
electronically,
and high-accuracy timing signals must be recorded together with the data,
so that the signals obtained at different antennas can be accurately
synchronised during correlation. In practice, the amplitudes and, especially,
phases of the measured complex visibility data unavoidably contain 
unknown errors, which can conveniently be expressed via antenna-based
complex gain factors:
\begin{eqnarray*}
V^{obs}_{ij} & = & G_iG_J^*V^{true}_{ij}
\end{eqnarray*}
\noindent
where $V^{obs}_{ij}$ and $V^{true}_{ij}$ are the observed and true visibility
functions on the baseline between antennas $i$ and $j$, and $G_i$, $G_j$ are 
the complex gain factors for antennas $i$ and $j$. The complex gains $G$
must be determined and removed from the data in order
to achieve the highest-quality images possible for the radio-telescope
array used. This is normally done iteratively, via alternate application of
self-calibration (Fort and Yee 1976, Cotton 1979, Readhead \& Wilkinson 1978,
Readhead et al. 1980, Cornwell \& Wilkinson 1981) and a deconvolution method, 
such as CLEAN (H\"ogbom 1974). 

\section{Matching and Aligning Images at Different Frequencies}

Since AGN are variable, it goes without saying that multi-frequency data to 
be compared must correspond to epochs separated by time intervals appreciably 
less than the timescales for variability of the source.  
When preparing data for techniques involving comparison of multi-frequency data,
various instrumental differences between the datasets must also be
taken into account.  VLBI datasets at different frequencies will have 
different angular resolutions and sensitivities to structures on various
scales due to the different baseline coverages of the observations.  
One approach to reducing these differences when comparing multi-frequency images
is to match the baseline coverages at different frequencies by giving relatively
low weights to the longest baselines at higher frequencies and to the shortest 
baselines at lower frequencies, e.g., via tapering of the visibility data. 
Alternatively, the images can be obtained without such weighting, but then all be
convolved with the same CLEAN beam before comparison.  The size of the 
CLEAN beam to be
used in this case is ordinarily roughly equal to the central lobe of the dirty 
beam for the lowest frequency.  The
accuracy to which the positions of the CLEAN components are known is limited
by the resolution of the observing system, and convolving with a beam that was
much smaller than the central lobe of the dirty beam would result in a 
``superresolved'' image that may not be reliable.

Iterative imaging via self-calibration and a deconvolution algorithm such as 
CLEAN is generally quite effective, but the absolute information about the 
coordinates of the source on the sky is lost during phase self-calibration, 
which essentially places the centre of gravity of the radio brightness 
distribution at the phase centre, which has coordinates $(0,0)$. 
Because most radio-loud AGN are highly core-dominated, 
directly comparing multi-frequency self-calibrated VLBI images of AGN 
essentially amounts to aligning these images on the observed VLBI
core, which usually coincides very closely with the peak of the radio brightness
distribution.

However, the standard theory of extragalactic radio sources (e.g. 
Blandford \& K\"onigl 1979) predicts a frequency-dependent shift
in the location of the VLBI core due to opacity effects in the
core region.  Reabsorption of synchrotron radiation takes place
in the ultra-compact region near the central engine of an AGN, a
mechanism which is more efficient at low frequencies.
Consequently, the peak brightness appears further along the jet
axis in lower frequency observations.  Thus the alignment of
multi-frequency images on their VLBI core results in a
misalignment between images of different observing frequencies.
This prediction is supported by observation: the
frequency-dependent shift in core position has been measured for
several quasars and micro-quasars (see Lobanov 1998 and
references therein), and the effect is discussed in terms of its
impact on high precision astrometry by Fey (2000), Charlot (2002),
Ros (2005) and Boboltz (2006). Lobanov (1998) explains in detail how the
frequency-dependence of the shift depends on physical conditions
near the central engine.

It is thus necessary to correctly align images prior to applying
multi-frequency data-analysis techniques.  This can be achieved
in one of two ways.  The first is by phase-referenced
observations, first employed by Marcaide \& Shapiro (1984), in which a nearby
source (or sources) is observed along with the target source.
The reference source would ideally be a point source, to
eliminate structure effects including those discussed above, but
this is rarely possible since most sources show extended
structure on the milliarcsecond scales available with VLBI.  The
position of the target source relative to the reference source
can then be determined.  The
second method involves aligning images according to the positions of
optically thin jet components (i.e. components optically thin to synchrotron
radiation, so that their positions are not affected by absorption
effects such as those occuring in the core) that are present in both images
to be compared.  This can be non-trivial, particularly if the
source has a complicated structure, but has been employed
effectively by several authors, such as Paragi et al. (2000), who used
this method to determine the radio core shift in 1823+568.
This difficulty in aligning complex images without distinct
optically thin components detected at all frequencies to be
compared was the main motivation for us to consider alternative 
methods of image alignment.

\section{Image Alignment via Cross-Correlation}

The method we have developed to align multi-frequency images is based on the
cross-correlation technique widely used in many fields, including
biomedical signal processing and imaging (Panescu 1993, Frank \& McEwan
1992) and remote
sensing (Hartl 1976). Cross-correlation provides a measure of how closely
correlated (i.e. how alike) two functions are.  The use of this measure in
image alignment gives an objective, quantitative assessment of how well two
images are aligned, and does not depend on the presence of very compact 
features.  By applying different shifts between images and
calculating the cross-correlation coefficient for each shift, it is possible to
determine which shift results in the best alignment.  The normalised
cross-correlation coefficient we used (see, e.g., Dunn \& Clark 1974) is defined
in two-dimensions as:
\begin{equation}
r_{x y}=\frac{\sum_{i=1}^{n}\sum_{j=1}^{n}(I_{\nu 1,i j}-\overline{I_{\nu 1}}) (I_{\nu 2, i j}-\overline{I_{\nu 2}})}{\sqrt{\sum_{i=1}^{n}\sum_{j=1}^{n} (I_{\nu 1, i j}-\overline{I_{\nu 1}})^{2}\sum_{i=1}^{n}\sum_{j=1}^{n} (I_{\nu 2, i j}-\overline{I_{\nu 2}})^{2}}}
\label{corrcoef}
\end{equation}

\noindent where $n$ is the number of pixels in each direction in the 
two-dimensional images to be compared, $I_{\nu 1, ij}$ and $I_{\nu 2, ij}$ 
are the intensities for the maps at frequencies $\nu 1$ and $\nu 2$ at pixel
$(RA_i, Dec_j)$, and $\overline{I_{\nu 1}}$ and $\overline{I_{\nu 2}}$ are 
the mean values of these two intensities over the region analyzed.
Although the source emission at a given location varies with frequency, all 
the radiation observed at radio frequencies is believed to be synchrotron
radiation from the same population of relativistic electrons.  Therefore the
optically thin total-intensity ($I$) structures should be well-correlated 
despite
local changes in the spectrum of the synchrotron emission due to variations in 
the local magnetic field, interaction with the surrounding medium and other
effects.  Using this (reasonable) assumption, it is possible to determine
the shift to be applied between maps by comparing the structures
of optically thin regions of the source at different frequencies.  The
highest correlation between dual-frequency images should therefore be obtained
when the areas being compared correspond to the same physical region of the sky.
This method has the advantage that it takes account of the optically thin
emission from the \emph{entire} source, not just isolated compact components.

\section{Implementation of the Cross-Correlation Technique}

The most widely used software for the calibration, imaging and analysis of radio
interferometric data is the National Radio Astronomy Observatory AIPS 
(Astronomical Image Processing System) package. We have written
a C program to implement the cross-correlation technique, which is
external to but compatible with the NRAO AIPS package. 
The input to the program are two images in the format produced by the 
AIPS task IMTXT, and the program outputs files which can be
imported back into AIPS using the task FETCH.  IMTXT allows the user
to export an AIPS image as a text file containing an array of floats representing
the map values at each pixel location.  Conversely, FETCH imports a text file
in the format exported by IMTXT as a map file which can be displayed by
any AIPS tasks that work with images (e.g. KNTR, TVLOD).  

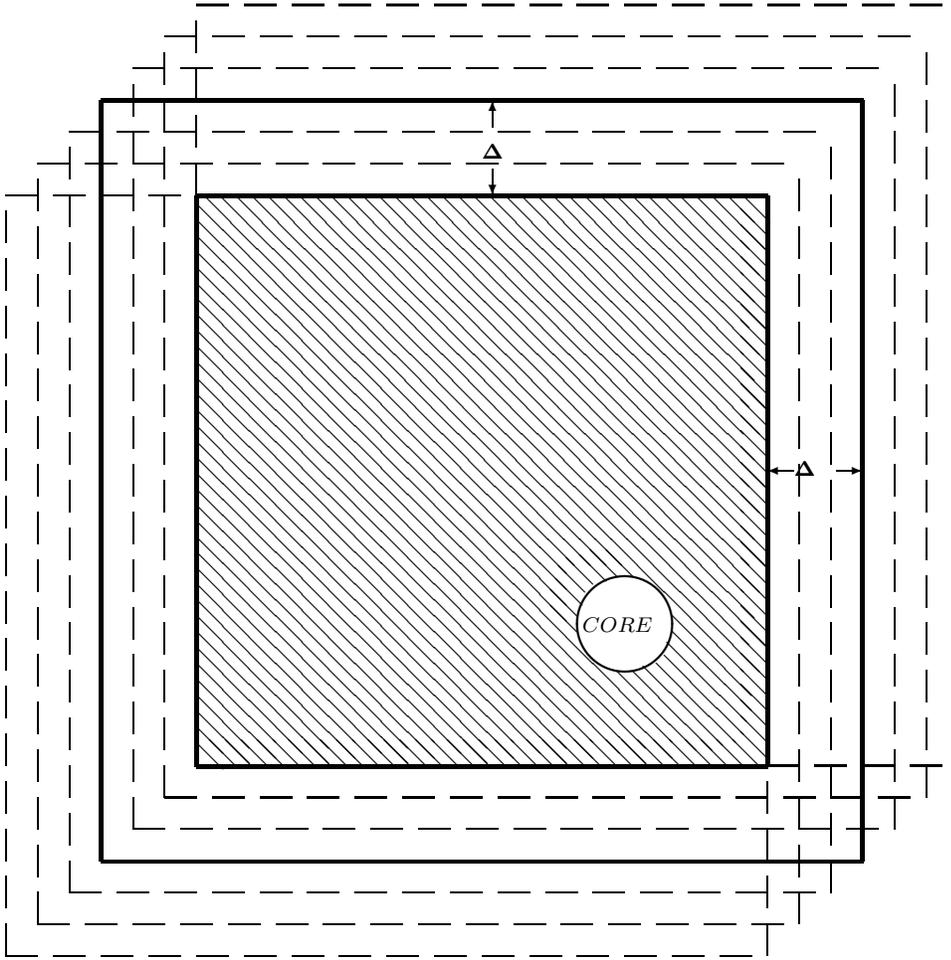
\begin{figure}
\begin{picture}(360,360)
\newsavebox{\map}
\savebox{\map}(288,288)[bl]{
\multiput(0,0)(0,288){2}{\multiput(0,0)(18,0){16}{\line(1,0){12}}}
\multiput(0,0)(288,0){2}{\multiput(0,0)(0,18){16}{\line(0,1){12}}}
}
\multiput(0,0)(12,12){7}{\usebox{\map}}
\linethickness{0.5mm}
\multiput(36,36)(0,288){2}{\line(1,0){288}}
\multiput(36,36)(288,0){2}{\line(0,1){288}}
\multiput(72,72)(0,216){2}{\line(1,0){216}}
\multiput(72,72)(216,0){2}{\line(0,1){216}}
\thicklines
\put(234,126){\circle{36}}
\thinlines
\multiput(72,285)(3,3){2}{\line(1,-1){147}}
\multiput(246,111)(3,3){2}{\line(1,-1){39}}
\multiput(72,279)(9,9){2}{\line(1,-1){146}}
\multiput(241,110)(9,9){2}{\line(1,-1){38}}
\multiput(72,273)(15,15){2}{\line(1,-1){144}}
\multiput(237,108)(15,15){2}{\line(1,-1){36}}
\multiput(72,267)(21,21){2}{\line(1,-1){144}}
\multiput(231,108)(21,21){2}{\line(1,-1){36}}
\multiput(72,261)(27,27){2}{\line(1,-1){189}}
\multiput(72,255)(33,33){2}{\line(1,-1){183}}
\multiput(72,249)(39,39){2}{\line(1,-1){177}}
\multiput(72,243)(45,45){2}{\line(1,-1){171}}
\multiput(72,237)(51,51){2}{\line(1,-1){165}}
\multiput(72,231)(57,57){2}{\line(1,-1){159}}
\multiput(72,225)(63,63){2}{\line(1,-1){153}}
\multiput(72,219)(69,69){2}{\line(1,-1){147}}
\multiput(72,213)(75,75){2}{\line(1,-1){141}}
\multiput(72,207)(81,81){2}{\line(1,-1){135}}
\multiput(72,201)(87,87){2}{\line(1,-1){129}}
\multiput(72,195)(93,93){2}{\line(1,-1){123}}
\multiput(72,189)(99,99){2}{\line(1,-1){117}}
\multiput(72,183)(105,105){2}{\line(1,-1){111}}
\multiput(72,177)(111,111){2}{\line(1,-1){105}}
\multiput(72,171)(117,117){2}{\line(1,-1){99}}
\multiput(72,165)(123,123){2}{\line(1,-1){93}}
\multiput(72,159)(129,129){2}{\line(1,-1){87}}
\multiput(72,153)(135,135){2}{\line(1,-1){81}}
\multiput(72,147)(141,141){2}{\line(1,-1){75}}
\multiput(72,141)(147,147){2}{\line(1,-1){69}}
\multiput(72,135)(153,153){2}{\line(1,-1){63}}
\multiput(72,129)(159,159){2}{\line(1,-1){57}}
\multiput(72,123)(165,165){2}{\line(1,-1){51}}
\multiput(72,117)(171,171){2}{\line(1,-1){45}}
\multiput(72,111)(177,177){2}{\line(1,-1){39}}
\multiput(72,105)(183,183){2}{\line(1,-1){33}}
\multiput(72,99)(189,189){2}{\line(1,-1){27}}
\multiput(72,93)(195,195){2}{\line(1,-1){21}}
\multiput(72,87)(201,201){2}{\line(1,-1){15}}
\multiput(72,81)(207,207){2}{\line(1,-1){10}}
\put(298,184){\vector(-1,0){10}}
\put(298,182){$\mathbf{\Delta }$}
\put(314,184){\vector(1,0){10}}
\put(184,298){\vector(0,-1){10}}
\put(180,302){$\mathbf{\Delta }$}
\put(184,314){\vector(0,1){10}}
\put(218,123){$CORE$}

\end{picture}
\caption{\small{Implementation of image alignment by cross-correlation. A
sub-area (shaded region) of the first map (outlined in bold) is compared with
the overlying region of the second map.  This sub-area remains constant,
while the corresponding region of the second map changes as it is shifted
relative to the first map (some possible shifted positions are outlined with
dashed lines).  The cross-correlation coefficient between the two regions,
given by equation (1), is computed each time to provide a measure
of how well the areas are correlated.  The method assumes that the highest
correlation is achieved when the areas being compared refer to the same
physical region of the sky. The circle labeled ``core'' represents a region
with a specified elliptical or circular shape (corresponding to the beam
shape) coincident with the position of the optically thick core, that is 
omitted from Map~1 during the cross-correlation calculations.}}
\label{corrfig}
\end{figure}

The input images to the program must have the same pixel size and
numbers of pixels (i.e. so that both the size of a pixel and the overall size
of the images correspond to the same area on the sky), convolved
with the same beam, and exported from AIPS using the task IMTXT.  The user
specifies the maximum trial shift to be applied between the images.  The
procedure used to calculate the shift that best aligns the input images is as
follows.

\begin{enumerate}
\item Strips of width $\Delta $, where $\Delta $ is the maximum shift to be
applied to the maps, are subtracted from each edge of the first map (Map 1).  An
area of the same shape as the restoring beam, but whose size is specified by
the user, is removed from the area in Map~1 surrounding the (at least partially)
optically thick core, whose position will be frequency-dependent and which 
usually corresponds to the peak of the map.  
In comparatively unusual cases, the brightest feature may not correspond to
the core; to enable the program to produce reliable results
for this situation, we have included the option of the user specifying the 
position of the core. 
\item The remaining area is used in the comparison (see Fig.
1).  The second map (Map 2) is shifted so that different regions 
overlay this area each time.  Since the area of Map~1 is
not changed, the normalised cross-correlation coefficients can be compared
directly to determine which part of Map~2 is best correlated with
the selected region.
\item Map~2 is first shifted so that its bottom left hand corner
corresponds to the bottom left hand corner of the selected sub-area of Map~1.
This corresponds to the maximum negative shift ($-\Delta , -\Delta $)
applied.  Map~2 is then shifted in right ascension, one pixel at a time, and
the cross-correlation coefficient computed each time, until the maximum
positive shift $\Delta $ is reached.  The image is then shifted by one pixel
in declination, and the correlation coefficients for the next row computed in
the same way.  This is repeated until the maximum shift in both directions
($\Delta , \Delta $) is reached, which occurs when the top right hand corner
of Map~2 corresponds to the top right hand corner of the selected subarea
of Map~1.
\item The program outputs to the screen the maximum correlation $r_{max}$
and the shift ($dRA$,$dDec$) at which it occurs.
\item Having found the best initial alignment between the two images, the 
program now applies the corresponding shift to Map 2, constructs a spectral-index
map and blanks any optically thick points in this map, taken to be  points with 
spectral indices $\alpha > 0$.  The theoretical limiting optically thick 
spectral index $\alpha = \frac{5}{2}$ is rarely reached, but it is generally 
accepted that a
positive spectral index implies some optically thick contribution in that
region.  The optically thick regions are blanked \emph{after} first calculating
the shift with only the core region blanked, since the shift between the maps
can result in a significant change in the spectral-index distribution, and
thus in those regions which have $\alpha > 0$.  The cross-correlation
procedure is then repeated, taking into account only the optically 
thin (i.e. unblanked) regions of the source, since the position of these regions 
is not affected by absorption effects.
\item The program again outputs to the screen the maximum correlation
$r_{max}$ and the shift ($dRA$,$dDec$) at which it occurs.  Positive 
shifts correspond to moving the second image downward (to the South) and to the 
left (East) relative to the first image.  In other words, a feature whose pixel
location is ($RA$,$Dec$) in the first map is located at 
($RA + dRA$,$Dec + dDec$) in the second map. 
\item Three files are output. One is simply a text file containing the 
array of cross-correlation values computed, while the other two are text files 
of the format recognised by the AIPS task FETCH.  One of these likewise
contains the cross-correlation coefficients; importing this image into AIPS and
plotting it (e.g with KNTR) shows the shape of the cross-correlation function.  
The other image text file displays the subarea of Map~1 that was compared; 
displaying this image can be useful in verifying the area to be
blanked around the optically thick core.  Ideally this area
should be large enough to cover virtually all of the core region, but not 
features in the inner jet; it should cover at least one beam size, and usually 
more, since the 
shape of the beam and the very high flux emanating from the core will dominate 
the VLBI $I$ structure here.
\end{enumerate}

In all cases we have tested, we found the
cross-correlation function to fall off monotonically from its peak.  This lends
credence to the hypothesis that the optically thin jet structures in the maps 
should be well correlated when aligned properly (if small-scale variations were
important the function could show secondary peaks where individual features
align well rather than falling off uniformly, but we have never found this
to be the case).  

Obviously, the accuracy of the estimated shift between a pair of images 
will depend to some extent on the pixel size in the images being compared.
In practice, it may be expedient, within reason, to use images with 
slightly smaller pixel sizes in order to derive more refined shift estimates. 
The results should not depend on whether the higher-frequency or 
lower-frequency images is used as Map~1: in either case, it is the 
relative shift that is being determined, and the shifts obtained with the 
higher-frequency map as Map~1 will be the negative of those obtained with
the lower-frequency map as Map~1.  In practice, it may be expedient to
compare results obtained blanking out core areas of several different sizes
(e.g. 2.0 times the beam area, 2.5 times the beam area etc.), to ensure 
that the optically thick core region is fully blanked, while as much of 
the optically thin inner jet is included in the correlation analysis 
as possible.

\section{Testing/Illustration of the Cross-Correlation Method}

\begin{figure*}
\begin{center}
\includegraphics[width=8cm]{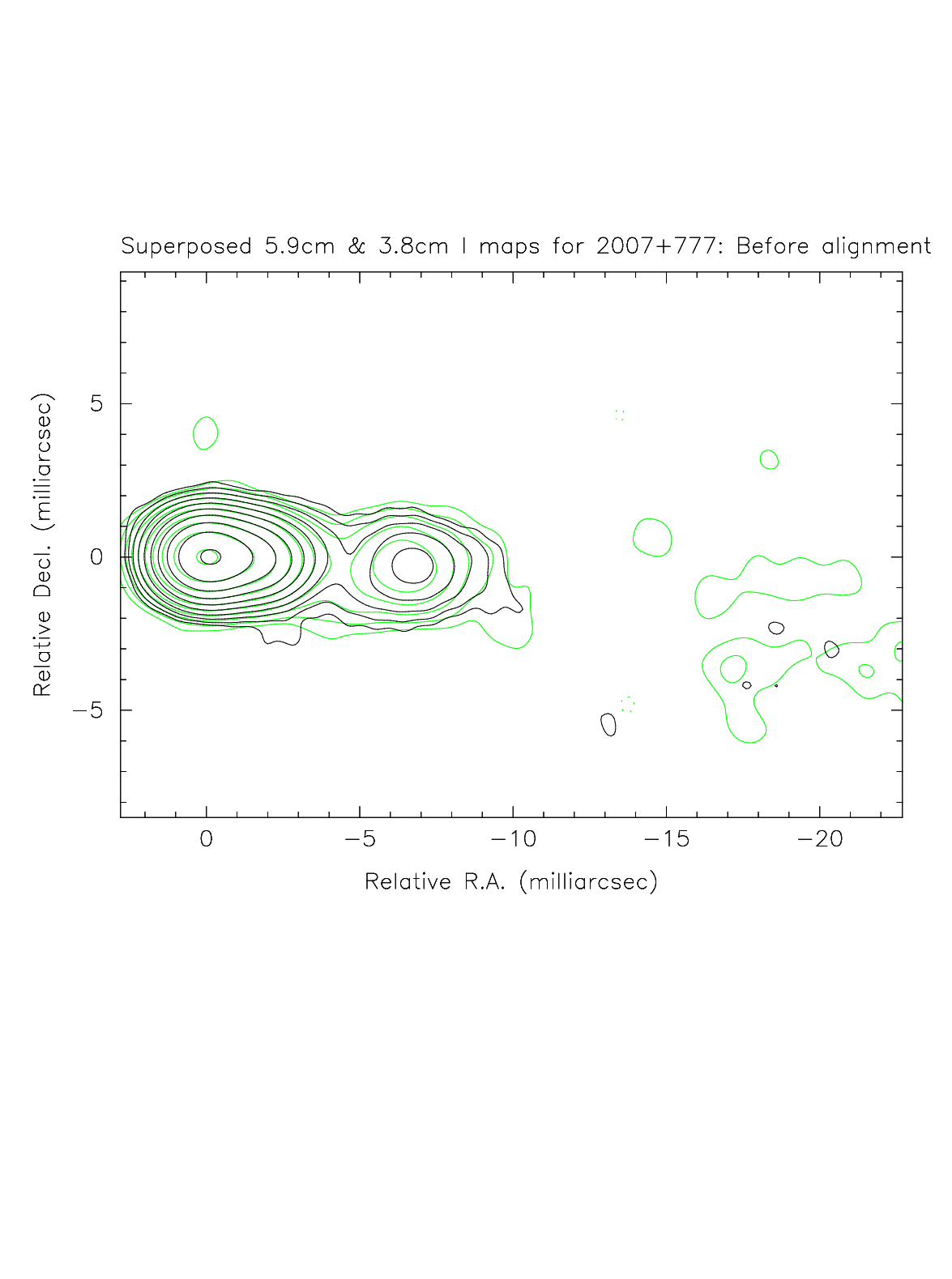}
\includegraphics[width=8cm]{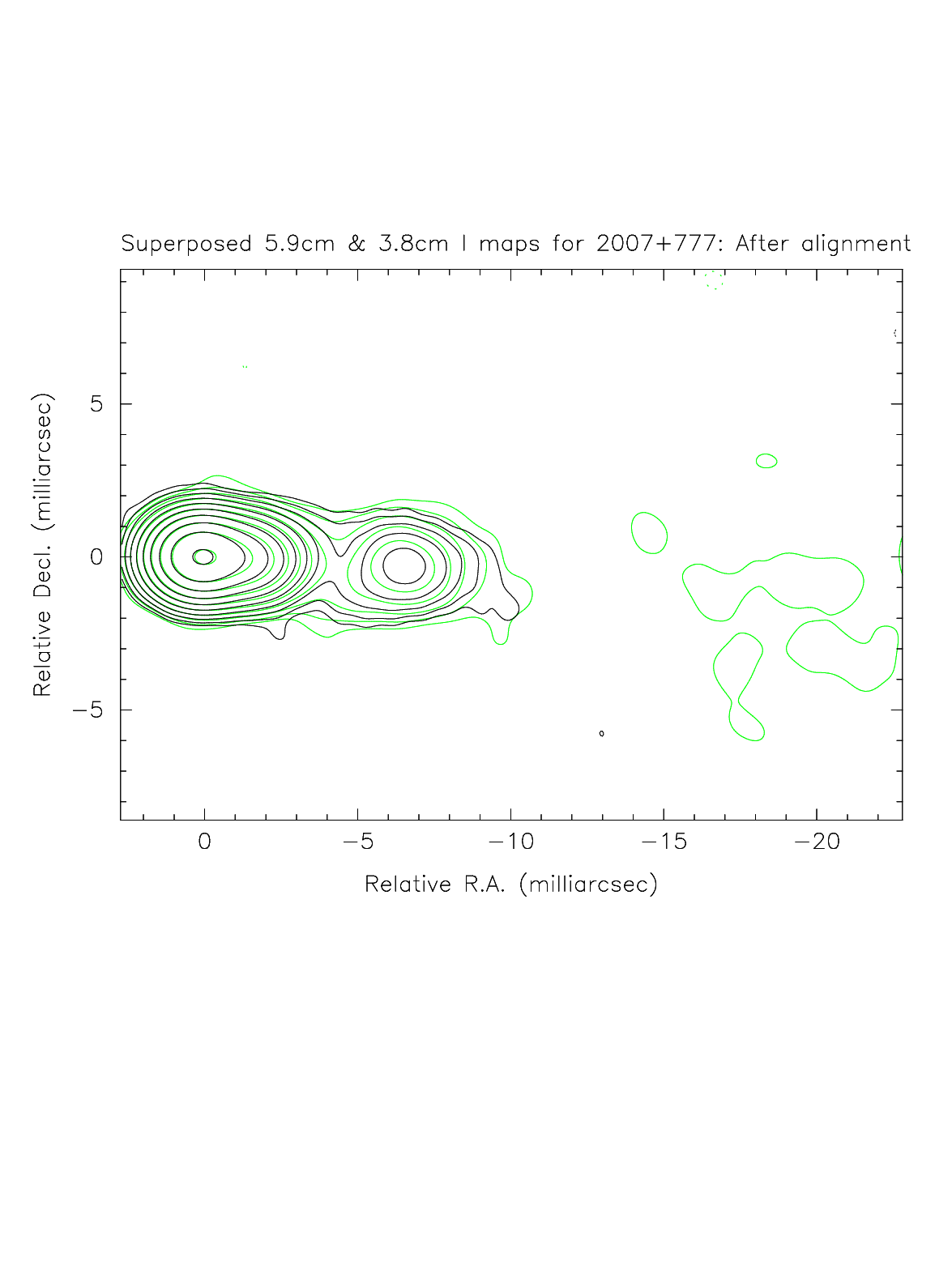}
\end{center}
\vspace*{-3.0cm}
\caption{\small{Maps of 2007+777 before (left) and after (right) alignment,
showing the $I$ contours at 5.1~GHz in green, with 
$I$ contours at 7.9~GHz superposed in black. The convolving beam was
$1.80\times 1.55$~mas in position angle $87.0^{\circ}$, and was the same 
in all cases. The bottom contour levels are 
$\pm0.20$, and the contours increase in steps of 1.98.  The peak brightnesses 
are 657~mJy/beam (7.9~GHz) and 547~mJy/beam (5.1~GHz).  The algorithm has 
clearly led to a very good alignment for the distinct, optically thin feature 
$\simeq 7$~mas from the core.}}
\label{2007_maps}
\end{figure*}

\begin{figure}
\begin{center}
\includegraphics[width=80truemm,angle=-90]{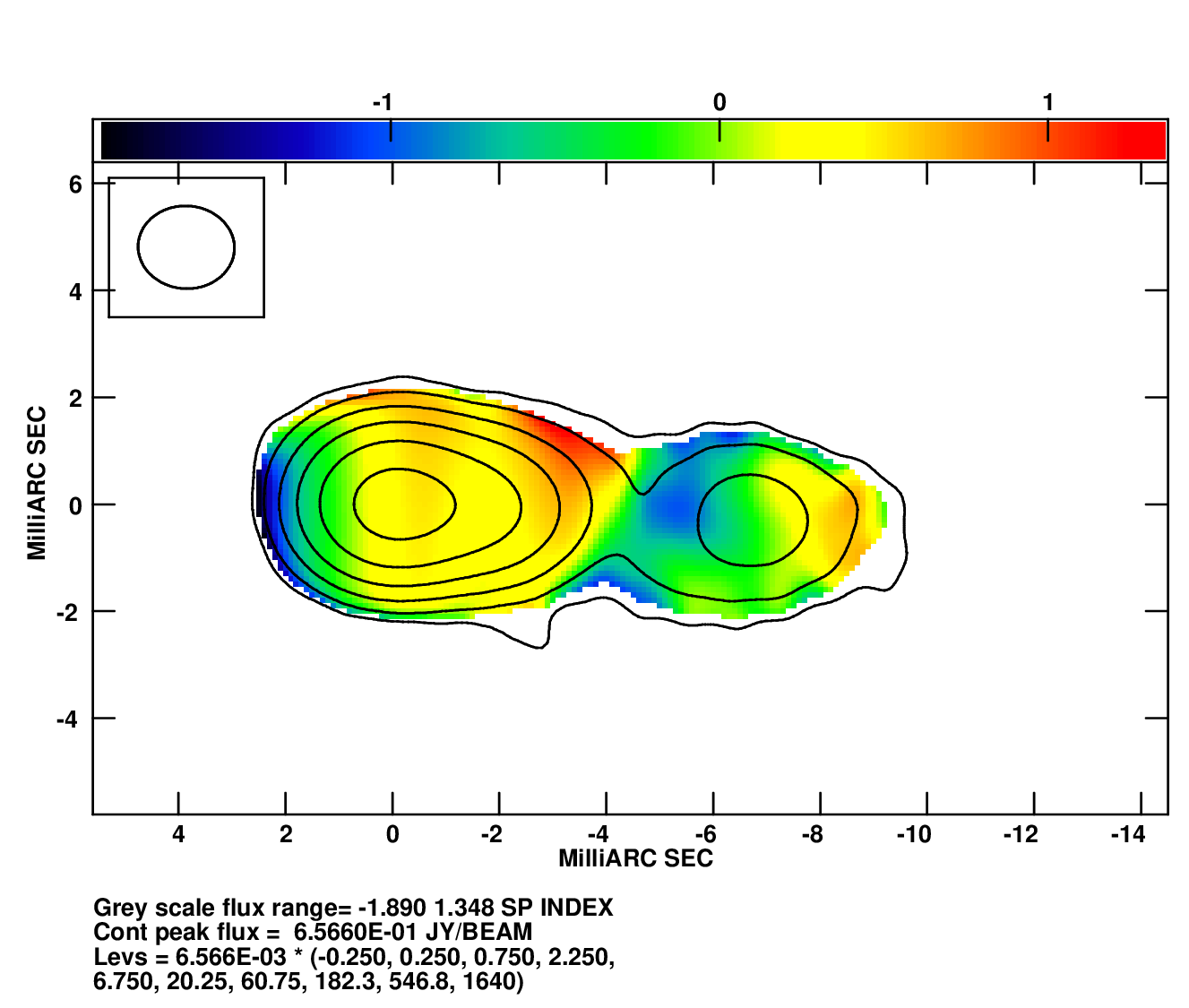}
\includegraphics[width=75truemm]{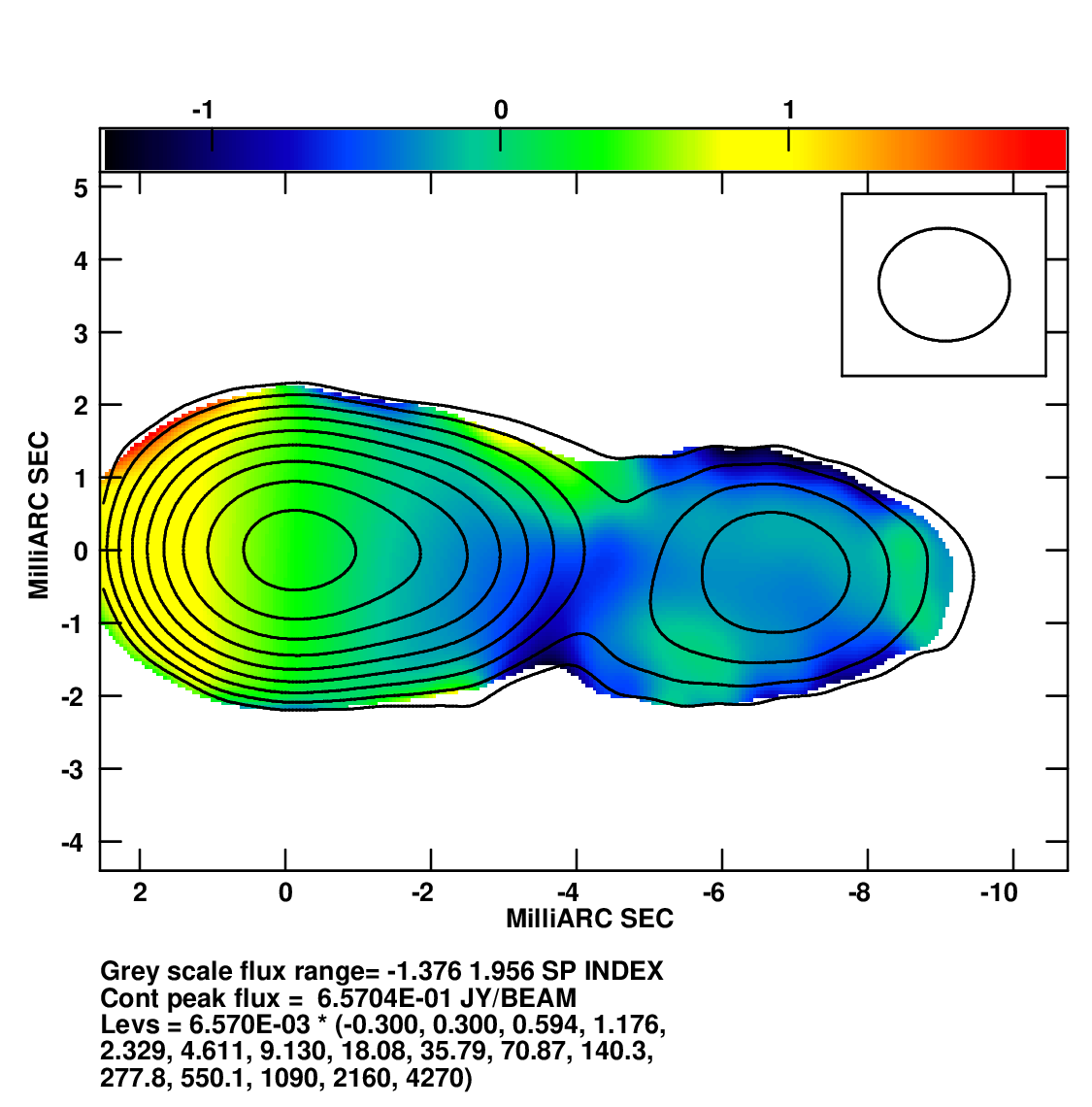}
\end{center}
\vspace*{-0.5cm}
\caption{\small{Colour spectral-index maps of 2007+777 before (top) and after 
(bottom) alignment, with the $I$ contours at 7.9~GHz from Fig.~2
superposed.  The convolving beam is shown in an upper corner of each image.
The shown ranges of spectral indices are from $-1.89$ to $+1.35$
(top) and from $-1.38$ to $+1.96$ (bottom).
Spectral artefacts due to misalignment are clearly visible in the 
top spectral-index map (false optically thin emission to the East of the $I$
peak and false optically thick emission at the Western end of the jet), 
which are absent from the 
bottom spectral-index map.}}
\label{2007_spixmaps}
\end{figure}

We present two examples of applying the program in practice to align
total-intensity images of AGN obtained with the NRAO Very Long Baseline
Array at 7.9~GHz and 5.1~GHz: (i) for 2007+777, which displays a fairly 
distinct optically thin jet feature that could be used to align the two maps in
the ``traditional'' way, and (ii) for 2200+420, which displays only fairly
amorphous optically thin jet emission.  The maps compared have matched 
cell sizes, map sizes and beam parameters.  The VLBA observations were made over
24 hours in July 2006 in a snapshot mode, so that the baseline coverage 
obtained was spread out over the time the sources were visible with all or 
nearly all of the VLBA antennas. The data was obtained simultaneously at the
different frequencies. The preliminary calibration and imaging of these data 
were carried out in AIPS using standard techniques; some initial results
are presented by O'Sullivan \& Gabuzda (2008), and a more complete analysis
is in preparation.

For each of these two sources, we show 
a superposition of the 7.9~GHz and 5.1~GHz
$I$ maps before and after applying the derived alignment shift 
(Figs.~2 and 4), and the spectral-index maps obtained 
before and after applying the derived alignment shift (Figs.~3 and 5).
In both cases, we used a cell size of
0.1~mas during the initial total-intensity mapping, but used final maps
made with a cell size of 0.05~mas as input to the program, to improve slightly
the accuracy of the relative shifts obtained. 

In the case of 2007+777,
the derived shift of the 5.1~GHz relative to the 7.9~GHz image was 4 pixels
(0.20~mas) to the West and 0 pixels to the South, in the direction of the 
VLBI jet, as expected. The correctness of this shift is immediately obvious 
via a visual inspection of the superposed $I$ images (Fig.~2) and the 
spectral-index maps before and after applying this shift (Fig.~3); the 
algorithm has obviously aligned
a distinct optically thin jet component located $\simeq 7$~mas to the West
of the map peak. In the case of 2200+420, the calculated shift of the 5.1~GHz
relative to the 7.9~GHz image was 4 pixels (0.20~mas) to the South and 
0 pixels to the East, again in the direction of the VLBI jet, as expected. 
It is less straightforward to estimate
the correctness of this shift directly from the superposed $I$ maps in 
Fig.~4, but the spectral-index map after applying this shift shows appreciably
more regular behaviour, with a smooth gradient in the spectral index from
North of the core region to the jet extending nearly directly to the
South; in particular, a spurious region of optically thin emission to
the North of the peak has disappeared.
 
We also show the results of applying the cross-correlation
technique to 4.8 and 1.6-GHz images of the AGN 1803+784  
(Gabuzda \& Chernetskii 2003), as an example of the operation of the program
when applied to two images at more widely separated frequencies. 
A superposition of the 4.8~GHz and 1.6~GHz
$I$ maps before and after applying the derived alignment shift 
is shown in Fig.~6, and the spectral-index maps obtained 
before and after applying the derived alignment shift in Fig.~7. 
A cell size of 0.50~mas was used during the initial total-intensity 
mapping, but final maps with a cell size of 0.25~mas were used as input 
to the shift program.  The derived shift of the 1.6~GHz relative to the 
4.8~GHz image was 5 pixels (1.25~mas!) to the West and 1 pixel to the 
North, in the direction of the VLBI jet. Although examination of the
superposed $I$ images (Fig.~6) does not enable an unambiguous estimate
of the needed shift ``by eye,'' the correctness of the shift derived
by our program is immediately obvious via a visual inspection of the 
spectral-index maps before and after applying this shift (Fig.~7). At
first glance, the 
``slanting'' boundary between the regions of optically thick and thin 
emission near the core seems suspicious, but in fact, 
the 4.8-GHz VSOP space-VLBI image for this epoch shows that the VLBI
jet initially emerges to the Northwest (Gabuzda 1999), and this 
feature likely real (the implied spectral-index gradient is roughly 
perpendicular to the direction of the small-scale jet).

Finally, we show the two-dimensional plots of the cross-correlation
functions output by the program for 2007+777, 2200+420  and 1803+784 (Fig.~8).
The cross-correlation plots contain a single peak,
with a monotonic fall-off in the correlation coefficient with distance from
the derived optimal alignment shift.

\begin{figure*}
\begin{center}
\includegraphics[width=8cm]{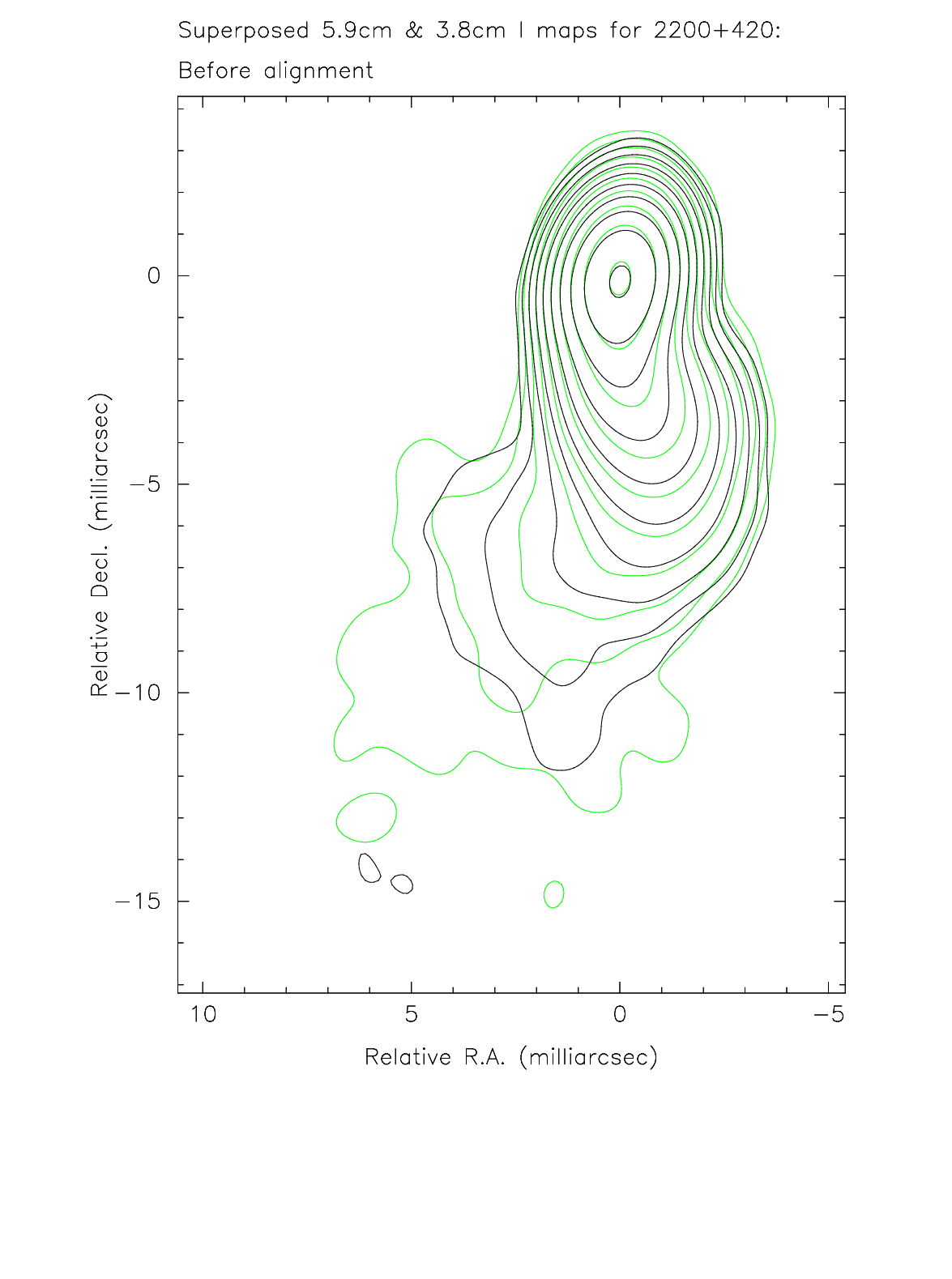}
\includegraphics[width=8cm]{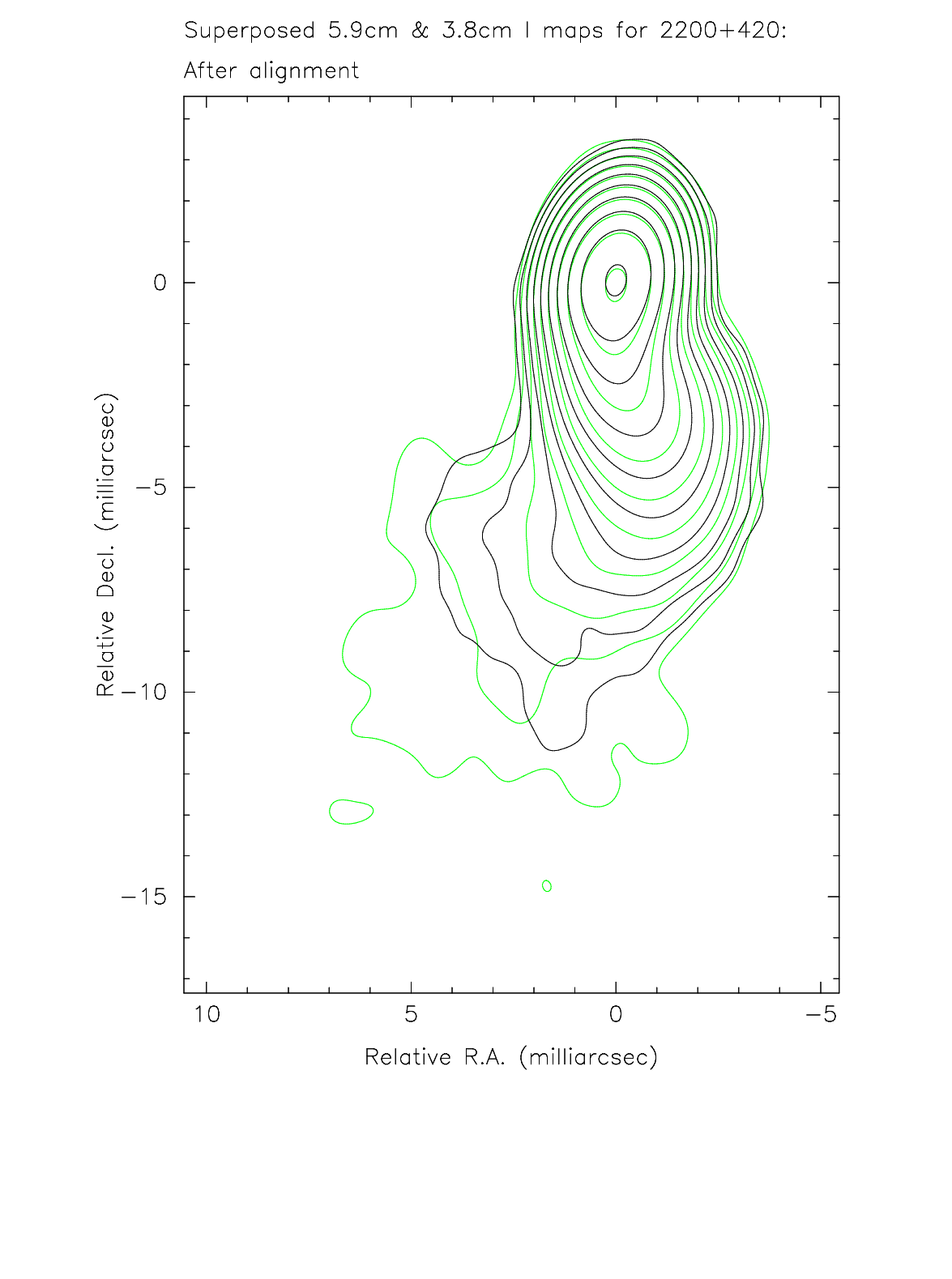}
\end{center}
\vspace*{-2.0cm}
\caption{\small{Maps of 2200+420 before (left) and after (right) alignment,
showing $I$ contours at 5.1~GHz in green, with $I$ contours 
at 7.9~GHz superposed in black. The convolving beam was
$2.25\times 1.59$~mas in position angle $-13.8^{\circ}$, and was the same
in all cases.  The bottom contour levels are
$\pm0.20$, and the contours increase in steps of 1.98.  The peak brightnesses
are 1940~mJy/beam (7.9~GHz) and 1520~mJy/beam (5.1~GHz). In this case, there 
is no
compact, distinct optically thin feature on which to base the derived shift
between the two images, but the algoritm has nevertheless properly aligned
the images, as is clear from a comparison of the corresponding spectral-index
maps in Fig.~5.}}
\label{2200_maps}
\end{figure*}

\begin{figure}
\begin{center}
\includegraphics[width=75truemm]{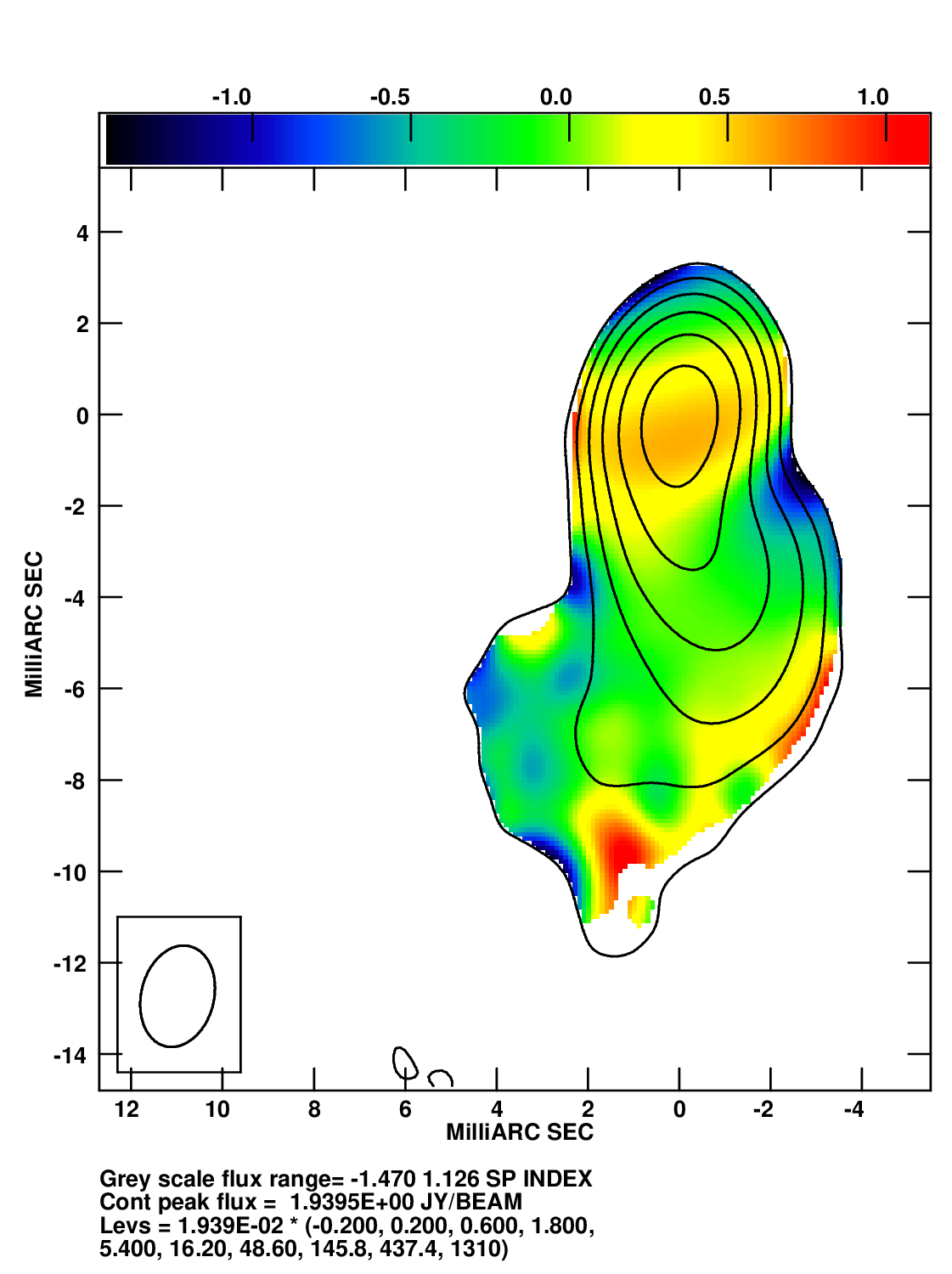}
\includegraphics[width=75truemm]{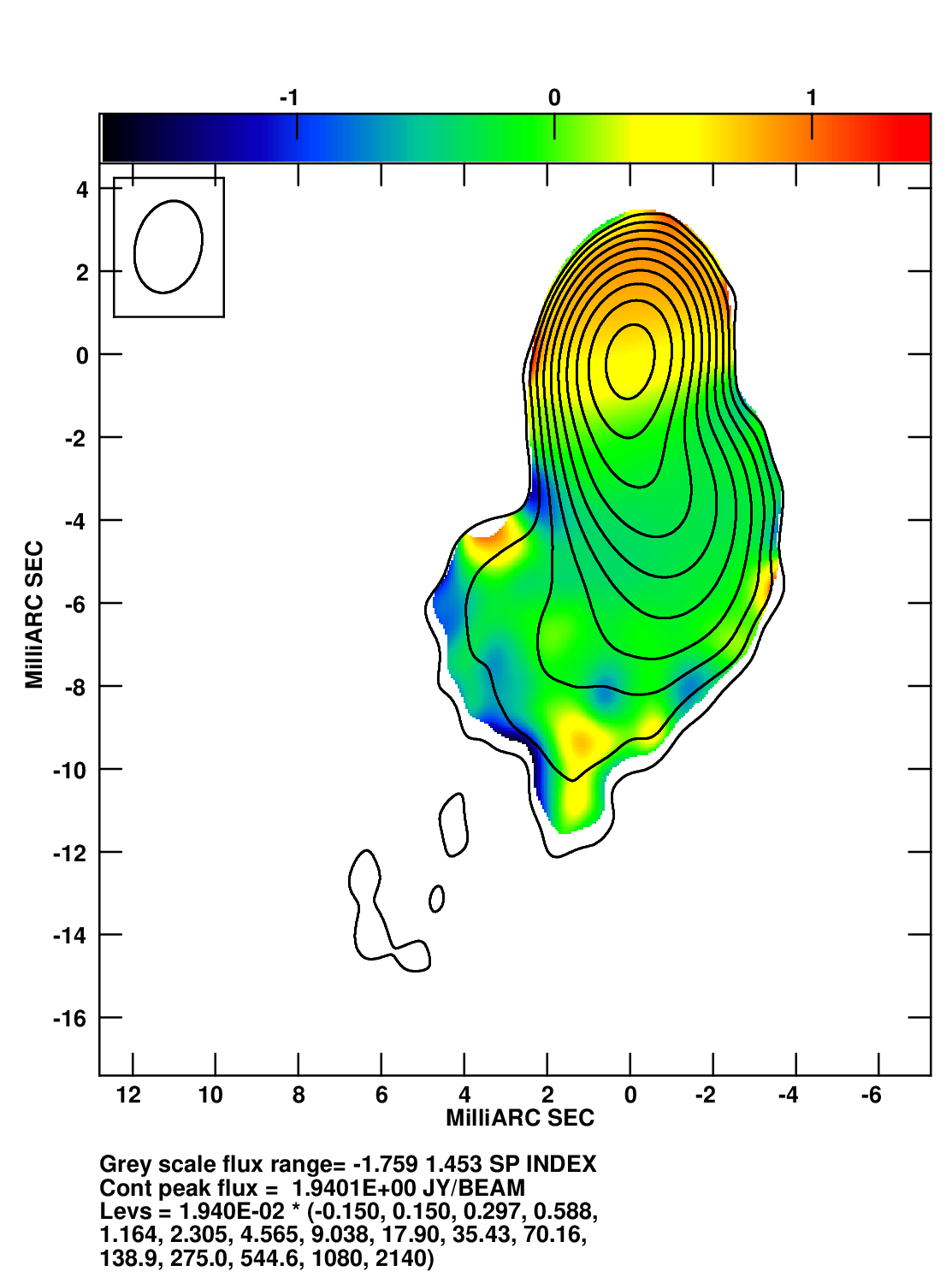}
\end{center}
\caption{\small{Spectral-index maps of 2200+420 before (left) and after 
(right) alignment, with the contours of total intensity at 7.9~GHz from
Fig.~4 superposed.  The convolving beam is shown in an upper corner of 
each image.  The shown ranges of spectral indices are from $-1.47$ to 
$1.13$ (left) and from $-1.76$ to $+1.45$ (right). 
False optically thin emission is visible to the North of the $I$ peak in the
top spectral-image map. This artefact
is absent from the bottom spectral-index map, and the range of spectral indices
in the VLBI jet is more moderate.}}
\label{2200_spixmaps}
\end{figure}

\begin{figure*}
\begin{center}
\includegraphics[width=8cm]{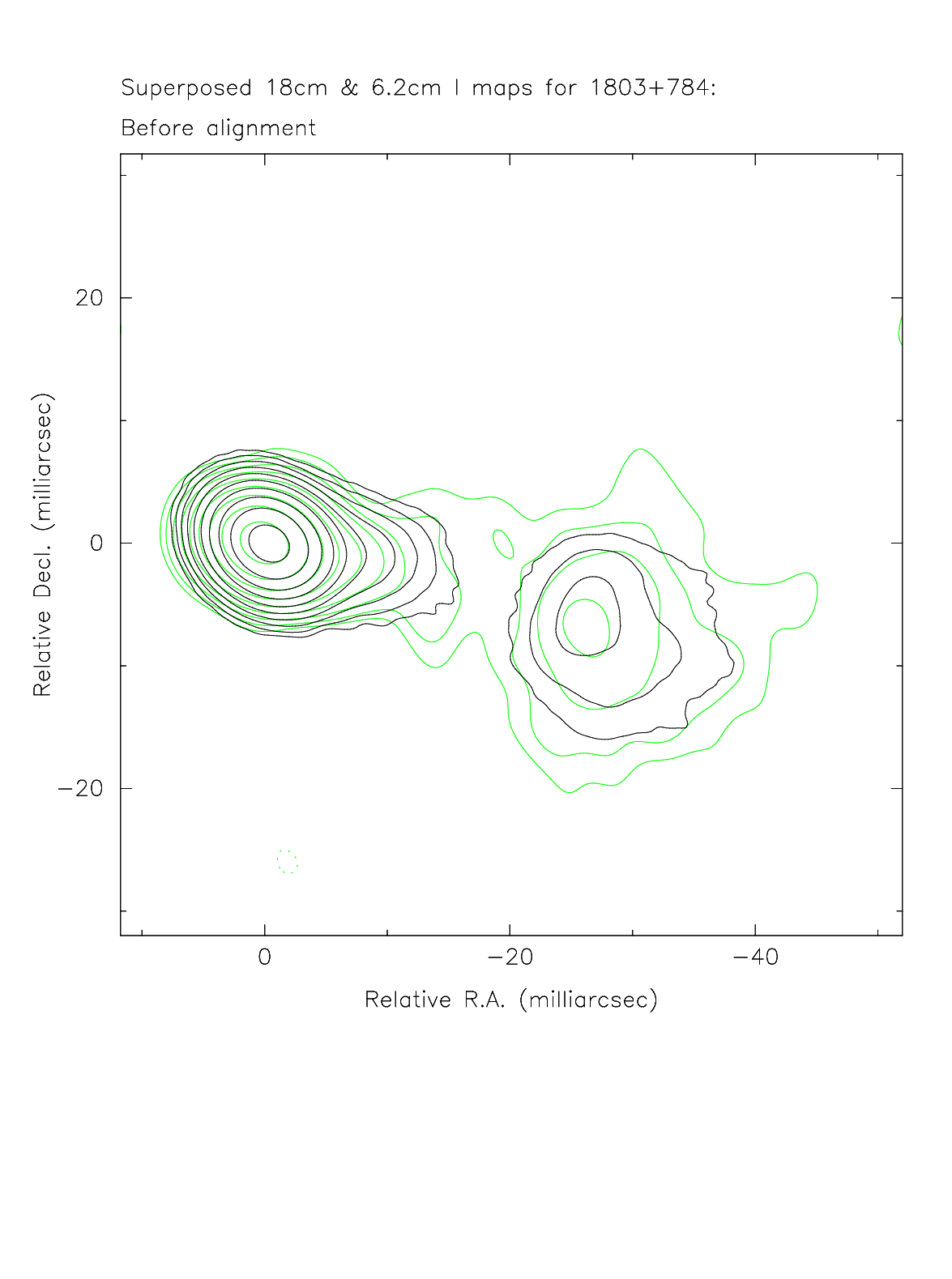}
\includegraphics[width=8cm]{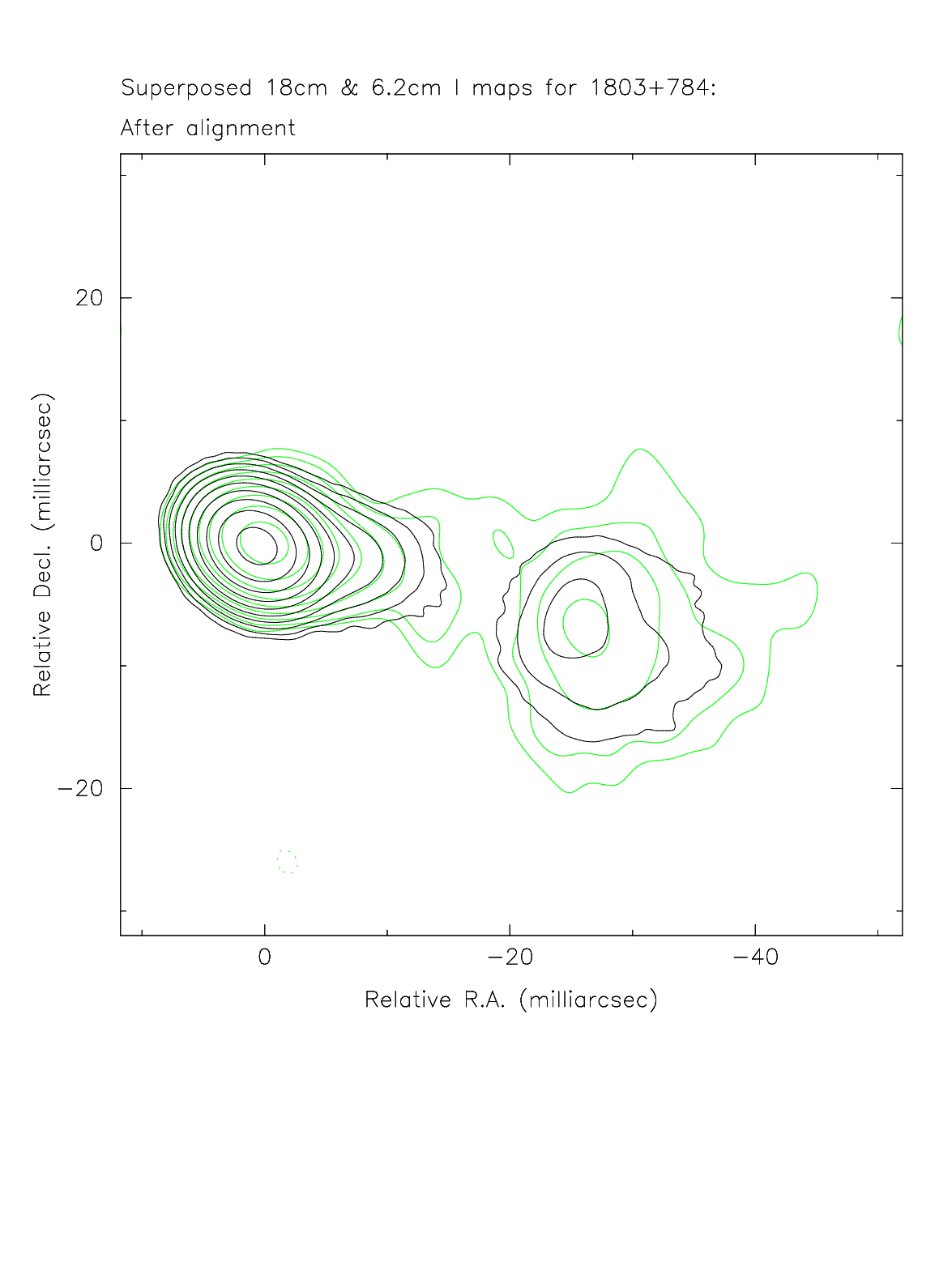}
\end{center}
\vspace*{-2.0cm}
\caption{\small{Maps of 1803+784 before (left) and after (right) alignment,
showing $I$ contours at 1.6~GHz in green, with $I$ contours 
at 4.8~GHz superposed in black as an example of the operation of the
cross-correlation technique applied to images separated by a larger
frequency difference. The convolving beam for both images was
$5.49\times 4.49$~mas in position angle $51.2^{\circ}$. 
The bottom contour levels are
$\pm0.15$ (4.8~GHz) and $\pm 0.28$~mJy (1.6~GHz), with the contours 
increasing in steps of 2.0 in both cases.  The peak brightnesses
are 2270~mJy/beam (4.8~GHz) and 1640~mJy/beam (1.6~GHz). The 4.8 and
1.6~GHz ``peaks'' of the somewhat diffuse jet feature roughly 25~mas from
the core are at appreciably different positions in the aligned maps, 
but an inspection of the spectral-index maps in Fig.~7 demonstrates that
the algoritmm has properly aligned the overall optically thin jet
structure. (Maps adapted from Gabuzda \& Chernetskii 2003.)}}
\label{1803_maps}
\end{figure*}

\begin{figure}
\begin{center}
\includegraphics[width=75truemm]{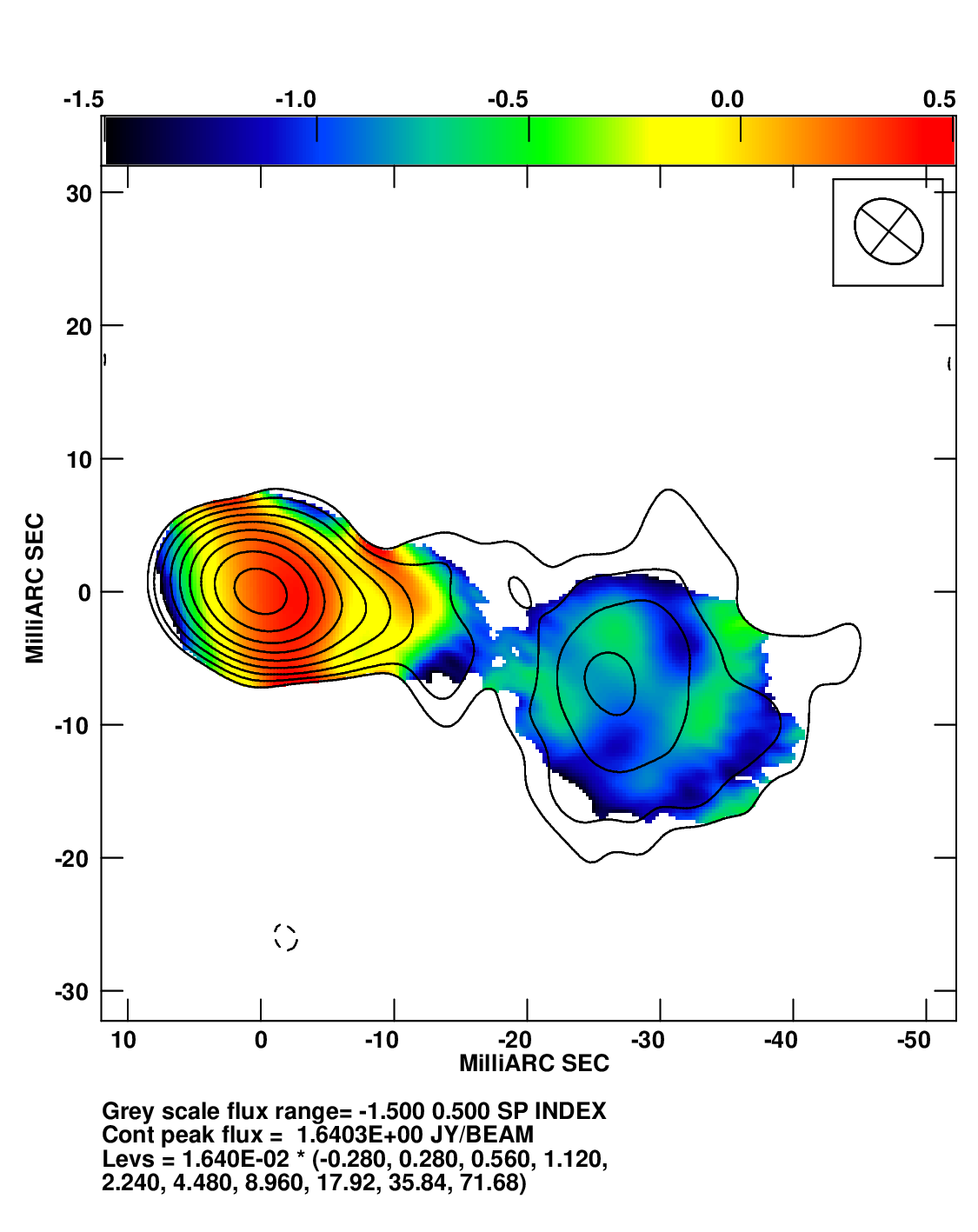}
\includegraphics[width=75truemm]{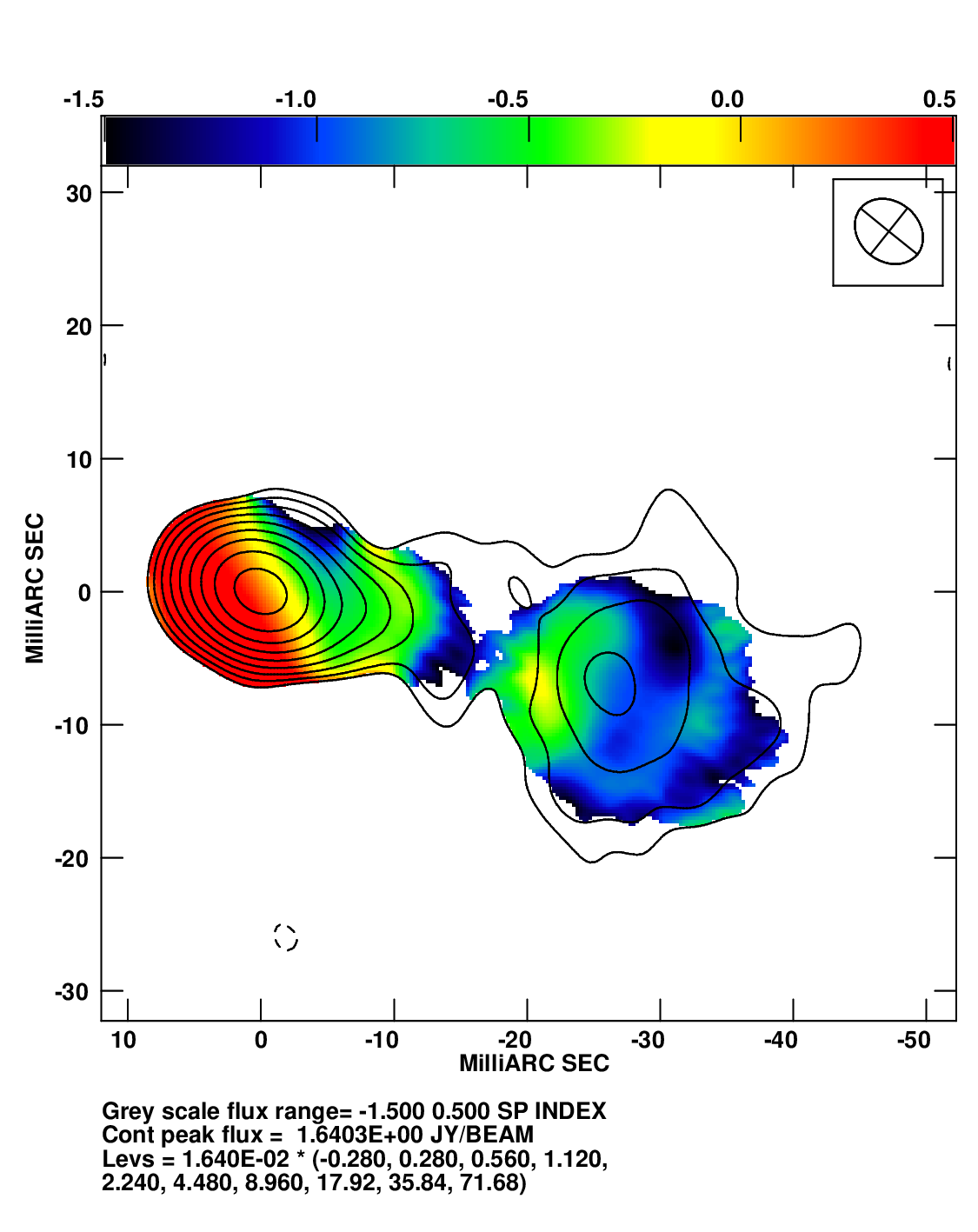}
\end{center}
\caption{\small{Spectral-index maps of 1803+784 before (left) and after 
(right) alignment, with the contours of total intensity at 1.6~GHz from
Fig.~6 superposed.  The convolving beam is shown in an upper corner of 
each image.  The shown ranges of spectral indices are from $-1.50$ to 
$0.50$ in both cases.  False optically thin emission is visible to the 
East of the $I$ peak in the top spectral-image map. This artefact
is absent from the bottom spectral-index map; the ``slanting'' boundary
between the regions of optically thick and thin emission near the core
seems suspicious, but in fact, the initial direction of the 4.8-GHz jet
on small scales is to the Northwest (Gabuzda 1999).}}
\label{1803_spixmaps}
\end{figure}

\begin{figure}
\begin{center}
\includegraphics[width=75truemm]{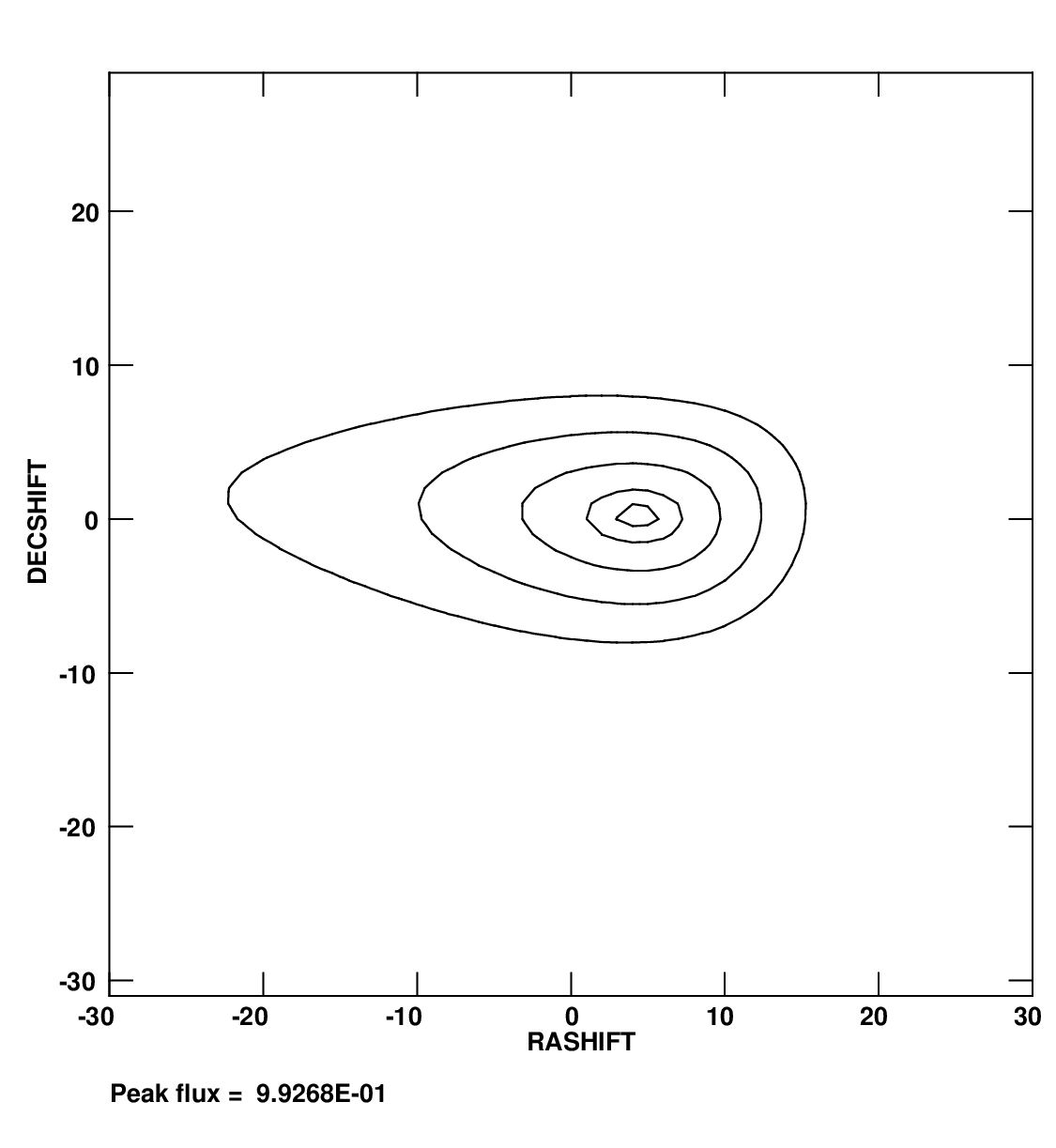}
\includegraphics[width=75truemm]{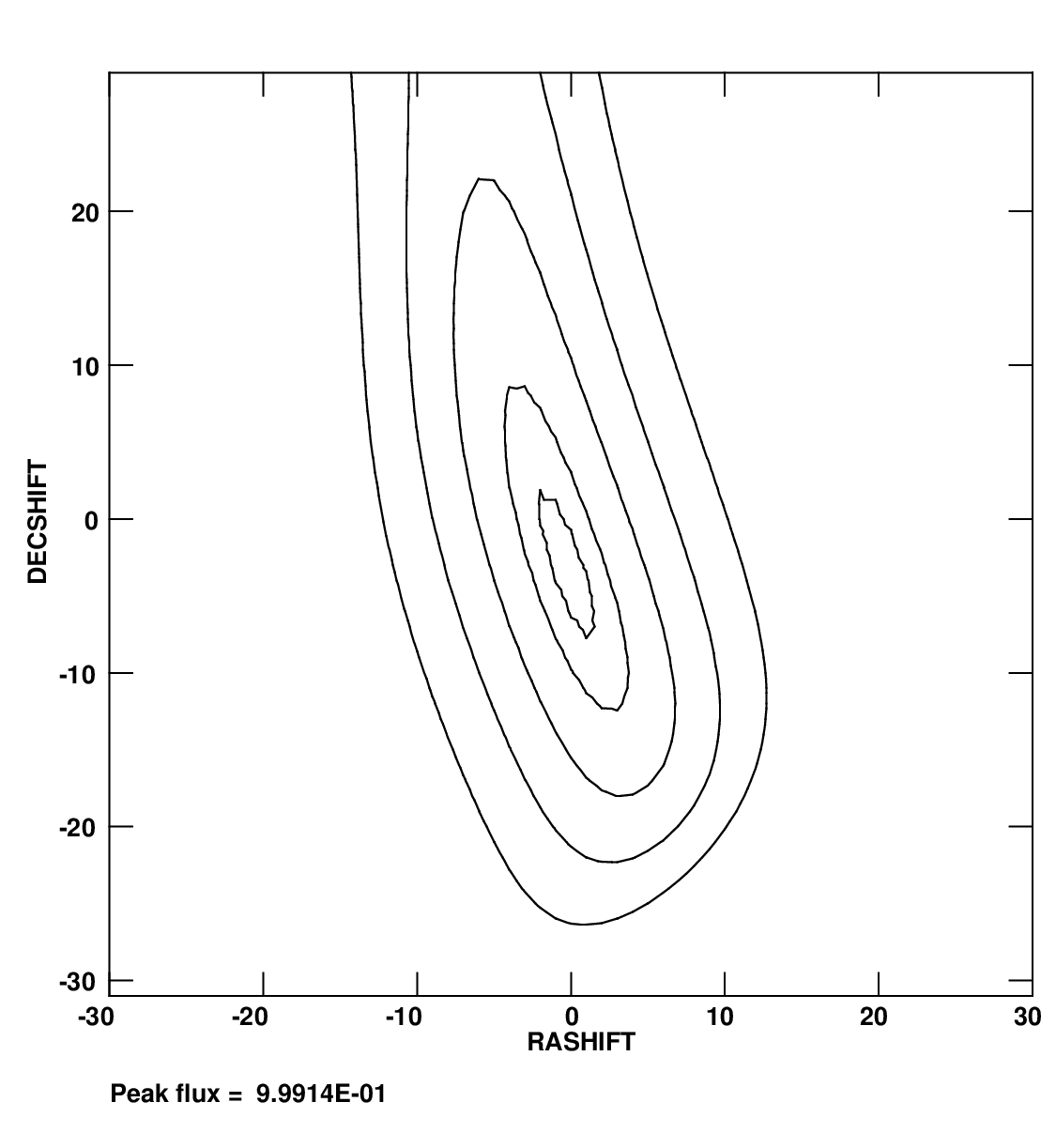}
\includegraphics[width=75truemm]{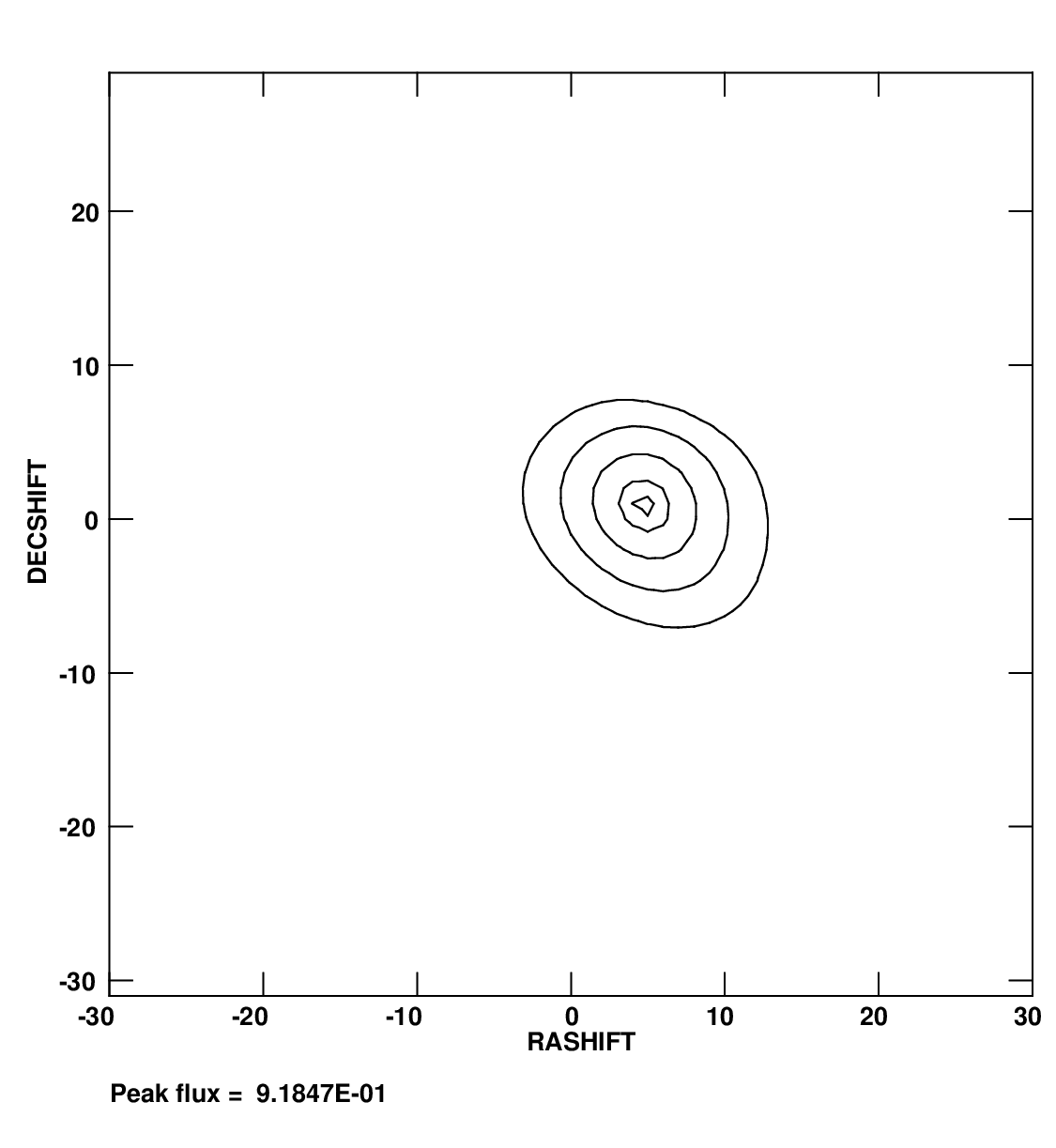}
\caption{\small{Cross-correlation functions obtained when the 7.9 \& 5.1-GHz
$I$ images for 2007+777 (top left) and 2200+420 (top right) and
4.8 and 1.6-GHz images for 1803+784 (bottom) were compared.  
Contour levels are 90, 95, 98, 99.5, and 99.9\% of the peak values of 
0.9927 (top right), 0.9991 (top left) and 0.9185 (bottom). The units 
plotted along the axes are pixels.}}
\label{corrplot}
\end{center}
\end{figure}

\section{Conclusion}

We have developed a C program to determine the shift between two VLBI images 
based
on a cross-correlation analysis of the images. In the past, it has been 
necessary to determine such shifts for images of AGN by aligning compact 
optically thin jet components, which can be a somewhat subjective and not
entirely unambiguous procedure. In addition, this method is difficult to
apply to complex and/or extended AGN jets without compact optically thin
features suitable for such an analysis. The great advantage of our new approach 
is that it provides a straightforward, objective means to determine the shift
between two images that makes use of all optically thin regions in the source
structure, not just individual chosen features. Our tests have shown that the
program produces reliable shifts for images both with and without distinct
optically thin features.  The code can be compiled using a standard C compiler,
and has been designed to take input files written by the AIPS task IMTXT, and 
to output files that can be read by the AIPS task FETCH, making it 
straightforward for radio astronomers familiar with AIPS to implement the 
code. The code can be
obtained by contacting D. C. Gabuzda (gabuzda@phys.ucc.ie).

\section{Acknowledgements}

We thank Shane P. O'Sullivan for allowing us to use unpublished images of
2007+777 and 2200+420 as examples of our alignment program, and for help
in making the spectral index maps for 1803+784. SC thanks Ger
Croke for helpful discussions on the cross-correlation technique. We
are also grateful to the referee, Bob Campbell, for helpful comments
that led to improvement of this paper.

\end{document}